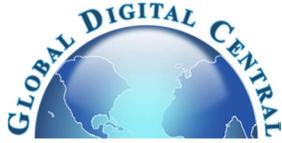
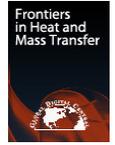

**Frontiers in Heat and Mass Transfer**

Available at www.ThermalFluidsCentral.org

# MULTICOMPONENT GAS-PARTICLE FLOW AND HEAT/MASS TRANSFER INDUCED BY A LOCALIZED LASER IRRADIATION ON A URETHANE-COATED STAINLESS STEEL SUBSTRATE

Nazia Afrin [a], Yijin Mao [a], Yuwen Zhang [a,*], J. K. Chen [a], Robin Ritter [b], Alan Lampson [b], Jonathan Stohs [c]

[a] Department of Mechanical and Aerospace Engineering, University of Missouri, Columbia, MO 65201, USA
[b] Tau Technologies, LLC, 1601 Randolph Rd., SE, 110s, Albuquerque, NM 87106, USA
[c] Air Force Research Lab-RDLT, 3550 Aberdeen Ave SE, Bldg 760, Kirtland AFB, NM 87117, USA

**ABSTRACT**

A three-dimensional numerical simulation is conducted for a complex process in a laser-material system, which involves heat and mass transfer in a compressible gaseous phase and chemical reaction during laser irradiation on a urethane paint coated on a stainless steel substrate. A finite volume method (FVM) with a co-located grid mesh that discretizes the entire computational domain is employed to simulate the heating process. The results show that when the top surface of the paint reaches a threshold temperature of 560 K, the polyurethane starts to decompose through chemical reaction. As a result, combustion products $CO_2$, $H_2O$ and $NO_2$ are produced and chromium (III) oxide, which serves as pigment in the paint, is ejected as solid parcels from the paint into the gaseous domain. Variations of temperature, density and velocity at the center of the laser irradiation spot, and the concentrations of reaction reactant/products in the gaseous phase are presented and discussed, by comparing six scenarios with different laser powers ranging from 2.5 kW to 15 kW with an increment of 2.5 kW.

## 1. INTRODUCTION

Because of the unique characteristics of coherency, hmonochromaticity and collimation, lasers have been widely used in various areas, such as etching and ablation of polyimide (Brannon, 1985, Sobehart, 1993), ablation of biological tissues (Lane et al., 1987; Lecartentier et al., 1993), and interaction with composite materials (Chen et al., 1995), to name a few. For many laser applications, scientists and engineers frequently encounter a situation that requires to couple multi-scales and multi-physics in solution of laser-material interaction. For example, laser cutting is one of the most important applications of laser in industry. To accomplish the task in terms of work quality and efficiency, a thoroughly understanding of the physics that are involved in the laser cutting process thus is of importance, including thermal transport across the object, change of material themophysical properties, phase change of melting and vaporization, chemical reaction in the material within or nearby the irradiated spot, and discretized particle ejection dynamics in the gaseous phase.

Literature survey indicate that numerous works on simulation of laser material processing has been published in the past decades. For example, Mazumder and Steen (1979) developed a three-dimensional (3D) heat transfer model for laser material processing with a moving Gaussian heat source using finite difference method. The results showed that some of the absorbed energy dismissed by radiation and convection from both the top and bottom surfaces of the substrate. Lipperd (2004) investigated laser ablation of polymers with designed materials to evaluate the mechanism of ablation. Zhou et al. (2007) developed a numerical model to simulate the coupled compressible gas flow and heat transfer in a micro-channel surrounded by solid media. Kim et al. (2001) studied the pulsed laser cutting using finite element method, and found that there were some fixed threshold values in the number of laser pulses and power in order to achieve the predetermined amount of material removal and the smoothness of crater shape. As a follow-up work, Kim (2005) further reported a 3D computational modeling of evaporative laser cutting process. Mahdukar et al. (2013) investigated laser paint removal with a continuous wave (CW) laser beam as well as repetitive pulses. The specific energy, a measure of the process efficiency that is defined as the amount of laser energy needed to remove per unit volume of paint prior to the onset of substrate damage, was found to be dependent of laser processing parameters. The result also showed that for a CW mode, the specific energy reduced with increase of laser scanning speed, irradiation time, and laser power.

The study of simultaneous fluid flow and heat and mass transfer in a coated medium induced by laser heating is scant. The objective of this work is to investigate the effects of laser irradiation on heat transfer and mass destruction of a urethane paint-coated substrate using a 3D numerical simulation. The paint starts to decompose through chemical reaction when the paint's hottest spot reaches a threshold temperature, 560 K. As a result, combustion products $CO_2$, $H_2O$ and $NO_2$ are produced and chromium (III) oxide, which is buried (as pigment) in the paint, is ejected as parcels from the paint into the surrounding gaseous phase. The results, including the variation of temperature and species concentration in the gaseous phase, amount of mass loss from the coated paint, and irradiation time before the onset of melting in the substrate steel, will be analyzed and discussed in detail.

* *Corresponding author. Email: zhangyu@missouri.edu*

## 2. PHYSICAL MODEL

The entire process of laser irradiation to a paint coated on a steel substrate includes: (1) thermal transport in the paint and substrate, (2) chemical reaction in the paint, and (3) heat and mass transfer of reactant and products in a multi-component gaseous phase.

### 2.1 Continuous Phase

For the gaseous phase, the governing equations of mass, momentum and energy are given as follows (Faghri et al., 2010):
(a) Continuity equation:
$$\partial \rho / \partial t + \nabla \cdot (\rho \mathbf{U}) = 0 \tag{1}$$
where $\rho$ and $\mathbf{U}$ are density and velocity of the gaseous phase, respectively. For a multi-component compressible flow, the mass density is component-weighted.
(b) Momentum equation:
$$\partial \rho \mathbf{U}/\partial t + \nabla \cdot (\rho \mathbf{U})\mathbf{U} = -\nabla p - \rho \mathbf{g} + \nabla \cdot \left[ \mu \nabla \mathbf{U} - 2\mu/3 \, trace(\nabla \mathbf{U})\mathbf{I} \right] + \mathbf{S_u} \tag{2}$$
where $p$ is pressure, $\mathbf{g}$ is gravitational acceleration, $\mu$ is dynamic viscosity, $\mathbf{I}$ is the unit matrix, and $\mathbf{S_u}$ is the source term that accounts for interaction between generated parcels and gaseous phase in each cell and it can be expressed as:
$$\mathbf{S} = -\sum_p N_p \mathbf{F}_p / \Delta V \tag{3}$$
where $N_p$ is the number of total particles in the p$^{th}$ parcel (see Section 2.2) that locates in one cell, $\mathbf{F}_p$ represents the force acting on a single particle in the parcel, and $\Delta V$ is the volume of the cell. The expression of $\mathbf{F}_p$ will be introduced later.
(c) Energy equation:
$$\frac{\partial}{\partial t}\left[\rho\left(h_s + \frac{1}{2}\mathbf{U}^2\right)\right] + \nabla \cdot \left[\rho \mathbf{U}\left(h_s + \frac{1}{2}\mathbf{U}^2\right)\right] = \nabla \cdot (\alpha h_s) - Dp/Dt + S_h \tag{4}$$
where $h_s$ is sensitive enthalpy, $\alpha$ is enthalpy-type thermal diffusivity, $Dp/Dt$ is material derivative of pressure $p$, and $S_h$ is the source term that represents heat transfer between particles and the surrounding gaseous phase. The mathematical expression of the source term can be written in the form of
$$S_h = -\sum_p N_p h_p / \Delta V \tag{5}$$
where $h_p$ is the enthalpy transferred between each individual particle in the p$^{th}$ parcel and the surrounding gaseous phase.
(d) Species concentration equation:
$$\partial \rho \omega_i / \partial t + \nabla \cdot (\rho \omega_i \mathbf{U}) = \rho \nabla \cdot (D_i \nabla \omega_i) \tag{6}$$
where $\rho$, $\omega_i$ and $D_i$ represent the mass density, mass fraction, and mass diffusion coefficient of specie $i$, respectively. The bulk density $\rho$ of the system is estimated through
$$\rho = p\psi \tag{7}$$
where $\psi$ is the bulk compressibility which is averaged over all gaseous species. Since all the gas species are considered as perfect gas, the bulk compressibility then can be estimated by
$$\psi = \sum_{i=1}^{N} \frac{\omega_i}{M_i} \Big/ RT \tag{8}$$
where $M_i$ denotes molar mass of specie $i$, $R$ is the gas universal gas constant, N is the number of species in the gaseous phase.

It is noted that the energy equation is solved in the enthalpy form, where temperature is implicit. However, it can be resolved by using the thermodynamic relationship between sensitive enthalpy and specific heat capacity (Bejan, 2006). And the specific heat capacity used is estimated through a forth order polynomial function of temperature (Gardiner, 2000). The temperature in the computational domain is thus corrected through iterative method according to the $h_s$-T diagram. In addition, the viscosity ($\mu_i$) of the i$^{th}$ specie is also treated as temperature dependent (Sutherland 1893). The bulk viscosity ($\mu$) and thermal diffusivity ($\alpha$) of the gaseous phase used in Eqs. (2) and (4) can be obtained by the molar-weight mean (Zhang and Faghri, 2000), which is similar to the bulk compressibility. The mass diffusivity of the i$^{th}$ specie ($D_i$) in the Eq. (6) can be determined by using the Maxwell-Stefan mass transport model (Bird et al., 1960) that considers the multi-species system as a special binary system:
$$D_i = (1 - X_i) \Big/ \sum_{j, j \neq i}(X_i / D_{ij}) \tag{9}$$
where $X_i$ denotes the molar fraction of specie $i$, and $D_{ij}$ are the mass diffusivity between species $i$ and $j$ that is temperature and pressure dependent, and the relationship is given by reference (Faghri 2010).

For the solid domain, it encompasses two layers: the bottom layer is stainless steel (AISI 304L) and the top layer is the paint that is a homogeneous mixture of binder (40%) and pigment (60%). Only heat conduction equation is solved for this domain. For the stainless steel, both thermal conductivity and heat capacity are temperature-dependent (Graves et al., 1991).

### 2.2 Chemical Reaction

The urethane based paint is considered in this work because it is an industrial standard. In the past two decades it has mostly replaced acrylic paints (Lu 2003). For this paint, the binder is polyurethane ($C_3H_7NO_2$) and the pigment is Chromium (III) oxide ($Cr_2O_3$). The overall chemical reaction occurring in the paint is "$4C_3H_7NO_2 + 19O_2 = 12CO_2 + 14H_2O + 4NO_2$."

The chemical reaction considered in this work is assumed as complete-type and zero order level. As mentioned previously, polyurethane will start decomposing when the temperature at the top surface of the paint reaches the threshold temperature. The produced gas species ($CO_2$, $H_2O$ and $NO_2$) then diffuse into the gaseous domain. At the same time, pigment (chromium (III) oxide) in the paint will be ejected into the gaseous domain from the paint surface. The reaction rate is approximated by Arrhenius equation (Mazumdar and Kar, 1995):
$$k_c = \left(RT/2\pi M_{O_2}\right)^{1/2} e^{-E/RT} \tag{10}$$
where $E$ is activation energy, and $M_{O2}$ is molar weight of Oxygen. Alternatively, the reaction rate can be defined as the slope of the concentration-time plot for a specie divided by the stoichiometric coefficient of that specie. For consistency, a negative sign is added if the specie is a reactant. Thus, the reaction rate constant can be represented as (Petrucci 2006):
$$k_c = -\Delta[C_3H_7NO_2]/\Delta t = -4\Delta[O_2]/19\Delta t = 4\Delta[CO_2]/12\Delta t$$
$$= 4\Delta[H_2O]/14\Delta t = 4\Delta[NO_2]/4\Delta t \tag{11}$$
Therefore, the reactant consuming rate and the products generating rate can be estimated through Eq. (11).

Another important concept that should be pointed out is parcel. Since it is computationally exhaustive to capture all particles' dynamic behaviors during the whole computational process due to the presence of high number of particles, a concept of parcel which is a collection of real particles (pigment in this case) is adopted to deal with solid particles dynamics in a fluid flow. In this sense, all the particles in one parcel share the same particle properties, i.e. size, velocity, temperature, etc. To facilitate the parcel generating mechanism, a field activation type of particle injection model that mimics the particle generation process is developed with the idea of introducing many local "injectors," Specifically, mass destruction is considered as parcel injection from "injectors" which are buried at the center of the cells that are adjacent to the coupled boundary between the solid and gaseous domains. The fields associated to those cells, namely temperature and oxygen concentration, determine whether injectors start or stop to work. In other words, when the temperature on the paint surface exceeds the threshold value and the oxygen concentration is sufficiently high at the same time, parcel will be generated and ejected into the gaseous domain according to the chemical reaction rate. The number of particles in single parcel generated in each time-step is estimated by
$$N = \Delta V_{rem} \rho^N = \Delta V_{rem} \frac{\omega_{pg} / \left(\rho_{pg} \frac{4}{3}\pi \bar{r}^3\right)}{\omega_{bd}/\rho_{bd} + \omega_{pg}/\rho_{pg}} \tag{12}$$

where $\Delta V_{rem}$ is the volume removed from an individual "injector," $\rho^N$ is number density of particles buried in paint, $\rho_{pm}$ and $\rho_{bd}$ are density of pigment and binder respectively, and $\bar{r}$ is the mean radius of particles in a parcel.

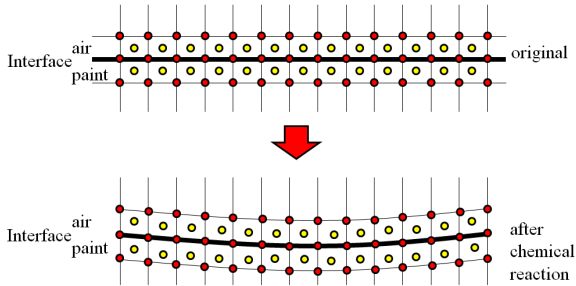

**Fig. 1** Illustration of the mesh deformation due to chemical reaction

According to the chemical reaction model, mass loss from paint can be determined; thus, the volume changed in the solid domain can also be obtained. In order to mimic the mass destruction, the mesh topology will be updated through moving the nodes shared by the solid and gaseous domain. In this work, it is proposed that the node displacement is calculated by using mass conservation at the solid/gas interface as follows:

$$d_{removal} = \left(\frac{4}{19} k_c C_{O_2} \Delta t\right) M_{bd} \Big/ \rho_{bd} \qquad (13)$$

where $k_c$ is chemical reaction rate constant, $C_{O2}$ is molar concentration of oxygen. The Fig. 1 is a two dimensional illustration that shows how the node displacements occur to mimic paint mass destruction during the chemical reaction. The yellow dots store temperature and oxygen concentration obtained by solving the corresponding equations; while the red dots store temperature and oxygen concentration interpolated from the yellow dots. As a consequence, the mass destruction can be more conveniently found through moving nodes based on the temperature and oxygen concentration at those red dots. In this case, the temperature in the solid domain and the oxygen concentration in the gaseous domain are interpolated to the corresponding nodes of its own grids. If the temperature at a node is higher than the threshold value and the oxygen concentration at that node is sufficiently high at the same time, then the nodes in both grids will be moved by a certain displacement that is determined through Eq. (13). After that, the concentration of reactants and products will be updated in both domains as well. Once the locations of the nodes in the paint region are updated, the nodes that share the same location but in the neighbor (gaseous) domain will be updated in order to make the domain spatially continuous. In addition, the internal nodes in both regions will also be updated to guarantee the mesh quality during computation. This can be done by solving the following 1-D diffusion equation for each domain,

$$\partial(\Delta Y)/\partial t = k_d \nabla^2 (\Delta Y) \qquad (14)$$

where $\Delta Y$ represent the moving distance in the y-direction in each time-step, $k_d$ is a diffusion coefficient that is equal to $1\times10^6$. These equations are solvable, because the boundary conditions are known according to the chemical reaction at each time step; it is also affordable due to its simplicity. By solving Eq. (14), the displacements of all the control volumes are known, and then a volume to point interpolation will be applied to obtain the displacement for each node.

### 2.2 Discretized Phase

In the aspect of parcels' motion, they can be determined by Newton's 2nd law. It should be noticed that the resultant should accounts for gravity and the drag force due to the velocity difference from the gaseous phase. The drag force is estimated in consideration of particle's sub-micro size (Ounis et al., 1991) as follows (Haider and Levenspiel, 1989):

$$\mathbf{F}_{drag} = \left[\frac{3}{4} m \mu_c \left(\frac{24}{C_c}\right) \Big/ \rho_p d_p^2\right] (\mathbf{U} - \mathbf{U}_p) \qquad (15)$$

where $\mu_c$ is molecular viscosity of the fluid, $\mathbf{U}_p$ is the velocity of the parcel, and $C_c$ is the Cunningham correction to Stokes' drag law (Ounis et al., 1991).

In addition, the simulation will be automatically terminated when the maximum temperature in the stainless metal reaches to its melting point (1,670 K) (Faghri et al., 2010), since the current work only focuses on the physical process of paint removal before the phase change of stainless steel occur, though the total simulation time and laser pulse irradiation time is set to be longer. The simulation is performed by using OpenFOAM-2.3.0 framework with the incorporation of our recently developed solver.

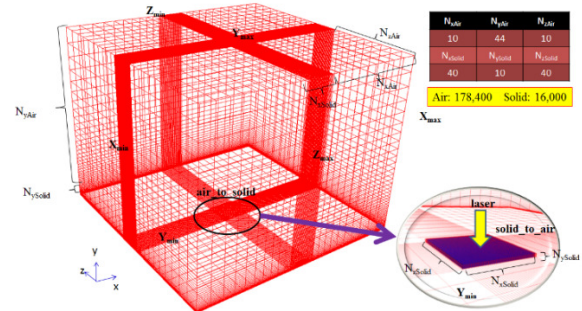

**Fig. 2** Illustration of mesh arrangement

### 3. RESULTS AND DISCUSSIONS

Figure 2 illustrates the physical and geometric model of the problem under consideration. A solid (paint + stainless steel) with a size of l×w×h (length×width×height) in the x-, z- and y-directions is placed at the bottom of the entire simulation domain that has a size of L×W×H (length×width×height). The dimensions used in the simulation are 250×250×200 mm³ for the gaseous domain and 30×30×1.35 mm³ for the solid domain. The thickness is 0.150 mm for paint and the rest 1.2 mm for stainless steel (AISI 304L). Six Gaussian laser beams with the same radius $r_o$ = 13.4 mm and laser power ranging from 2.5 kW to 15.0 kW with an increment of 2.5 kW are prepared. The material absorptivity is assumed to be constant, 0.8 for the paint and 0.1 for stainless steel (AISI 304L). The total laser irradiation time on the paint is set to be 10 s. The initial temperature of the entire domain is 300 K. As described previously, the mass diffusivities of all species are pressure and temperature dependent and their initial values can be found in the reference (Kreith, 2000). The specific heat capacity and the absolute viscosity of each species are given in the reference (Kreith, 2000). And the density, specific heat and thermal conductivity of stainless steel are available from the public accessible reference (Faghri et al., 2010). The properties of paint are estimated based on the volume-weighted average over all components. The threshold temperature for chemical reaction to take place is 560 K, the activation energy of the chemical reaction is 45 kJ/mol (Chang, 1994; Hentschel and Münstedt, 2001). The particles are assumed to have a mono-size distribution with a diameter of 750 nm. The initial velocity of each ejected parcel is zero. The initial flow is static with temperature of 300K, while the pressure is 1 atm. Mass concentration of oxygen and nitrogen are 21% and 79% respectively. On the boundary conditions, all the boundaries except the interface with solid region are considered as open to the surrounding air. For the solid region, the initial temperature is also 300K, while it is irradiated by a heat source, $2P\alpha_a e^{-2(x^2+z^2)/r_0^2} \Big/ \pi r_0^2$, which is considered as a surface source. The temperature is coupled with gaseous domain through the interface by satisfying energy conservation. A mesh independent study is first carried out before conducting the entire simulation in order to identify an optimal arrangement of the mesh. To overcome the difficulty from creating an eligible mesh configuration caused by the large spatial difference between the solid and gaseous domain, the entire computational domain is decomposed into 18 small blocks, and grading hexahedral cells are applied to each block with the purpose of reducing numerical diffusion caused by non-orthogonality, skewness and smoothness (Jasak).

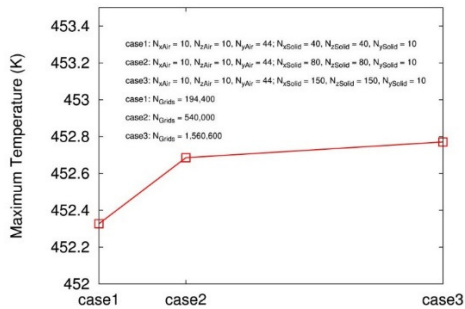

**Fig. 3** Maxmum temperatures in the paint vs three different mesh configurations

Fig. 3 shows the temperature variation after one time-step ($\Delta t = 1.25 \times 10^{-4}$ s) from three different meshes, with laser power of 15,000 watts. It is found that the maximum temperature does not show significant difference (< 0.02%) between the two finer meshes, 452.76 K vs. 452.68 K. The finest mesh that has a total of 194,400 cells, including 178,400 for the gaseous domain and 16,000 for the solid domain is adopted for the following simulations. The maximum temperature in the paint is chosen here as a benchmark because of its importance in affecting properties across the entire domain. Additionally, it plays a key role in activating chemical reaction.

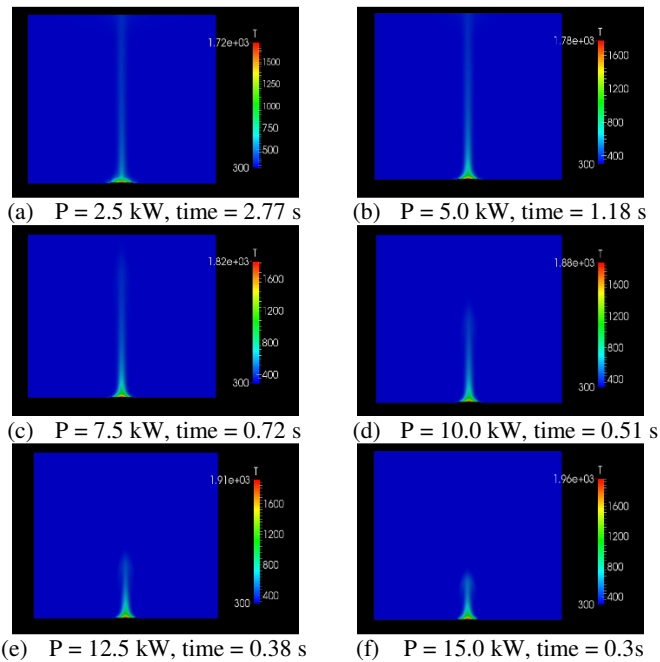

(a)  P = 2.5 kW, time = 2.77 s    (b)  P = 5.0 kW, time = 1.18 s

(c)  P = 7.5 kW, time = 0.72 s    (d)  P = 10.0 kW, time = 0.51 s

(e)  P = 12.5 kW, time = 0.38 s   (f)  P = 15.0 kW, time = 0.3s

**Fig. 4** Temperature distribution across the middle cross section area of the gaseous domain at the end of simulation

In this work, six scenarios with different laser powers are prepared to study how the laser power affects the chemical reaction rate and temperature, density, velocity and concentration of the resulting gas species from the heated paint. Fig. 4 shows the temperature distribution across the middle cross-section area of the gaseous domain at the end of simulation for all the six cases. The areas of hot region in the gaseous domain decrease with the increase in laser power. This trend is mainly attributed to the simulation time, 2.77 s, 1.18 s, 0.72 s, 0.51s, 0.38 s, and 0.30 s for the ascending laser powers. According to the absorbed laser powers and heating times, the total laser energies deposited into the entire system at the end of simulation are 6.925, 5.900, 5.400, 5.100, 4.750, and 4.500 kJ, respectively. The absorbed energy in the case of the lowest power is 1.53 times that of the highest power. The more the energy absorb, the more the gas species generate. In addition, a longer time for the species to move into the gaseous domain allows the hot species to spread over. Therefore, the lower the laser power is, the larger the area of hot region in the gaseous domain can be observed. It is also found from Fig. 4 that the maximum temperatures of the gas species are 1720 K, 1780 K, 1820 K, 1880 K, 1910 K, and 1960 K, respectively. The trend that the maximum temperature increases with laser power is because a shorter heating time not only limits the spread area of the hot species, but prevents the thermal energy in the hot region from diffusing into the surrounding colder area. As a consequence, a higher maximum temperature adjacent to the center of the heated spot is expected for the case with higher laser power.

Figure 5 reveals the time histories of temperature at the center of the laser heating spot on the top surface of paint. Apparently, a lower laser power leads to a lower maximum temperature and a longer simulation time, which is also confirmed by the results in the Fig. 4. Corresponding to the temperature distributions shown above, Fig. 6 shows the mass density distribution across the same cross-section area in the gaseous domain at the end of simulation.

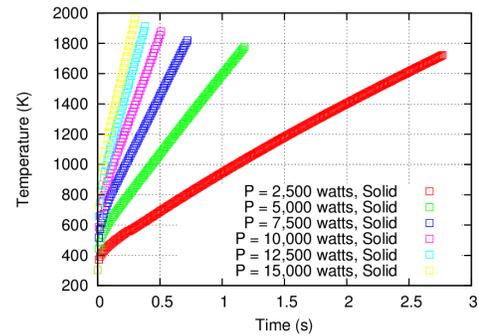

**Fig. 5** Time history of temperatures at the center of laser heating spot for the six laser powers

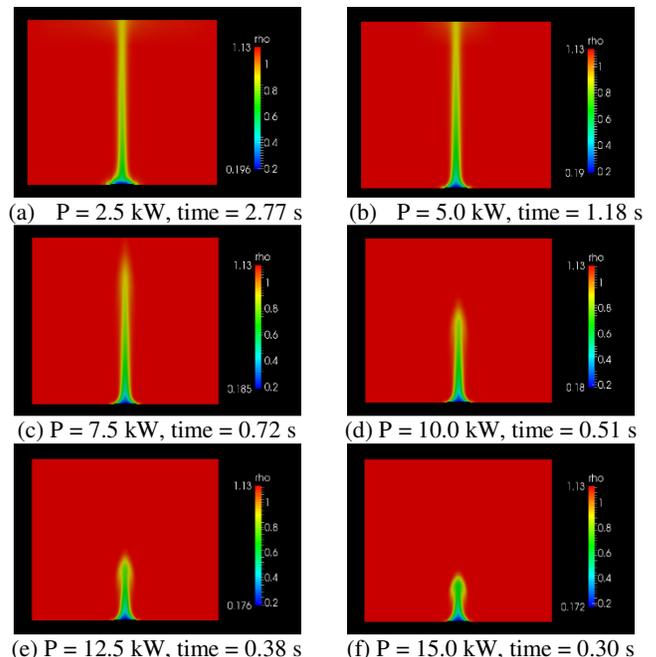

(a)  P = 2.5 kW, time = 2.77 s    (b)  P = 5.0 kW, time = 1.18 s

(c) P = 7.5 kW, time = 0.72 s     (d) P = 10.0 kW, time = 0.51 s

(e) P = 12.5 kW, time = 0.38 s    (f) P = 15.0 kW, time = 0.30 s

**Fig. 6** Density distributions across the middle cross-section area of the gaseous domain at the end of simulation

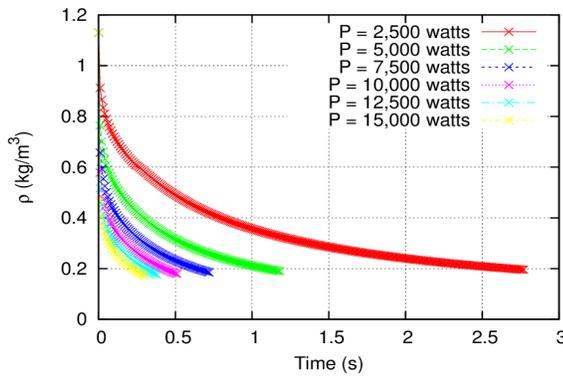

**Fig. 7** Density variations at the center of the laser irradiation spot with time

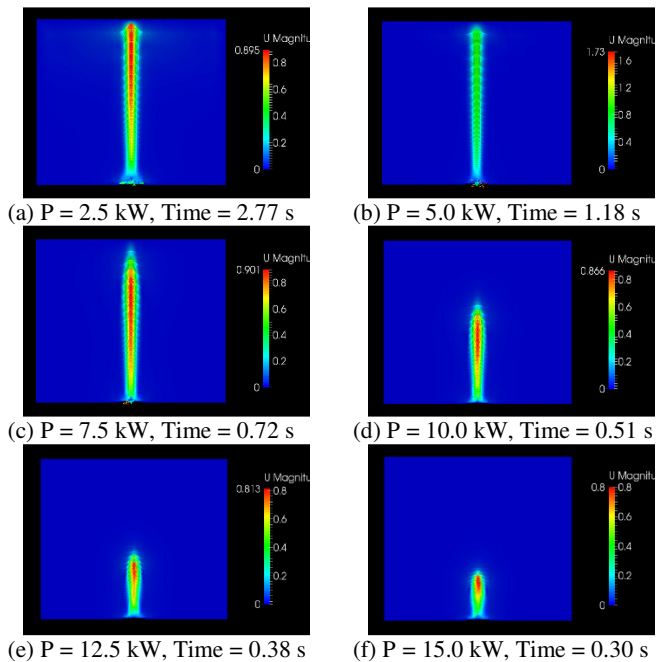

(a) P = 2.5 kW, Time = 2.77 s
(b) P = 5.0 kW, Time = 1.18 s
(c) P = 7.5 kW, Time = 0.72 s
(d) P = 10.0 kW, Time = 0.51 s
(e) P = 12.5 kW, Time = 0.38 s
(f) P = 15.0 kW, Time = 0.30 s

**Fig. 8** Velocity distributions across the middle cross section area of the gaseous domain at the end of simulation

Since the gas thermal state is determined by the ideal gas law which is a univalent function of temperature, it is found that the areas of lower mass density are similar as those distributions of temperature shown in Fig. 4. Accordingly, the higher the temperature is, the lower the density will be. As seen in Fig. 6, the minimum densities are 0.196, 0.190, 0.185, 0.180, 0.176, and 0.172 kg/m$^3$ for the six laser powers, respectively. The Fig. 7 presents the time histories of mass density of the gaseous phase close to the center of the laser heating spot. All the curves show a decreasing trend, falling from the initial value of 1.13 kg/m$^3$ at room temperature. Due to the fact that the chemical reaction rate is proportional to temperature, a higher laser power would result in a quicker decline of the gaseous mass density which is confirmed in Fig. 7.

Figure 8 plots the gas velocity distributions across the same cross-section area of the gaseous domain at the end of simulation. It is found that the maximum velocities are 0.898, 1.73, 0.901, 0.866, 0.813, and 0.800 m/s corresponding to the ascending laser powers at the end of the simulation. From the perspective of energy conservation, the more the laser energy deposited into the system, the more kinematic energy can be absorbed by the gaseous phase.

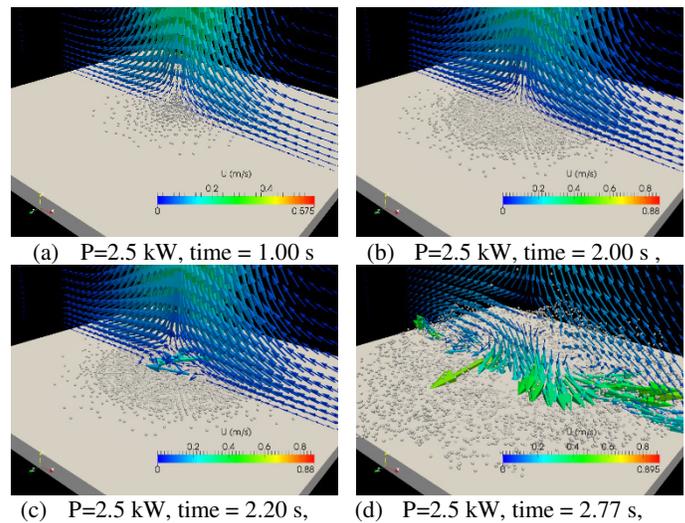

(a) P=2.5 kW, time = 1.00 s
(b) P=2.5 kW, time = 2.00 s,
(c) P=2.5 kW, time = 2.20 s,
(d) P=2.5 kW, time = 2.77 s,

**Fig. 9** State of parcel flow and gaseous phase at different times

The change of gas velocity against laser power confirmed this conclusion, except the case with laser power of 2.5 kW. In order to reveal more detailed results, a close look at the interaction between ejected parcels and gaseous phase, four snapshots, which are at time of 1.00 s, 2.00 s, 2.12 s and 2.77 s, are provided in Fig. 9. It can be seen that parcels (white dots) are lifted by the surrounding air as shown in Fig. 9 (a). And the entire velocity field is found symmetric-like. Figure 9 (b) shows the similar phenomena, but with more parcels and higher velocity. However, asymmetric velocity appear in Fig. 9 (c) due to the momentum transfer between parcels and gaseous phase which is the leading factor of affecting the entire velocity field in simulation domain. In addition, from the perspective of momentum conservation, the momentum exchange will slow down the gas velocity and accelerate the parcels' velocity if the drag force is larger than gravity. However, once the number of parcel is too huge, the momentum of gaseous phase will be completely drag down as a result of this intensive momentum exchange. In other words, the gas velocity will be pulled down toward the ground. The last snapshot, Fig. 9 (d) shows that the velocity at the zone that is close to the laser heating spot is directing to the ground. As a result, the parcels on the paint will be pulled to a place nearby as shown. A comparison between Fig. 9 (c) and (d) shows that the velocity does not have a significant increase during the period of 0.57s due to large number increase of parcels which certainly consume a large amount of momentum that hold by gaseous phase.

Figure 10 shows the mass concentration variations for the species, namely, $O_2$ as reactant and $H_2O$, $CO_2$ and $NO_2$ as products, adjacent to the center of the laser heating spot. It can be clearly seen in Fig. (a) - (f) that the concentration of the reactant $O_2$ keeps flat in the very beginning of heating process and then decreases, accompanying with the increase of the produced species. For those laser powers higher than 10 kW as shown in Fig. (g), (i) and (k), the concentrations of $O_2$ fall down very quickly once the lasers heat the paint. For all the cases, the decrease of $O_2$ concentration changes its course to increase at a turning point where the mass concentration is about 1.7%. The generation of the three product species depends upon the reaction rate and the mass diffusion whose intensity is governed by the concentration gradient produced by the continuous chemical reaction. For example, all the product concentrations increase with time as shown in Fig. (b), while the $H_2O$ concentration changes its trend from increase to decrease in latter period shown in Fig. (d) and (f) and remains almost no change in Fig. (h), (j) and (l). It is also observed that the increasing trends of all products can be categorized into the fast and slow zones. The fast zones appear immediately when the chemical reaction take place, and the slow zones come out later. In view of the fact that mass concentration adjacent to the center of the heating spot is contributed from the two competing mechanisms: chemical reaction and mass diffusion, the relative strength of the two parts can explain the trends shown in these figures. For the products, chemical reaction would increase the mass concentration, while mass diffusion tends to decrease it. Therefore, the fast zones suggest that the intensity of chemical reaction is relatively stronger than that of the mass diffusion, while the slow zones show the opposite.

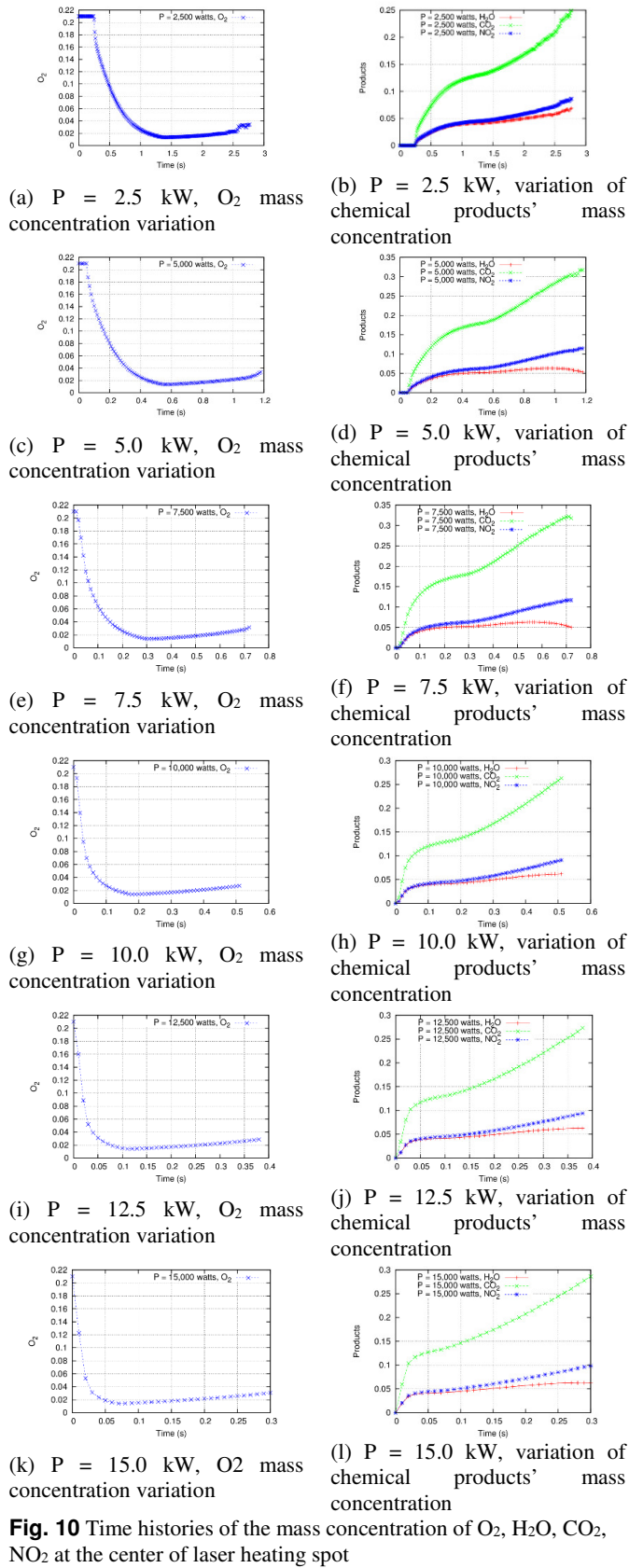

**Fig. 10** Time histories of the mass concentration of $O_2$, $H_2O$, $CO_2$, $NO_2$ at the center of laser heating spot

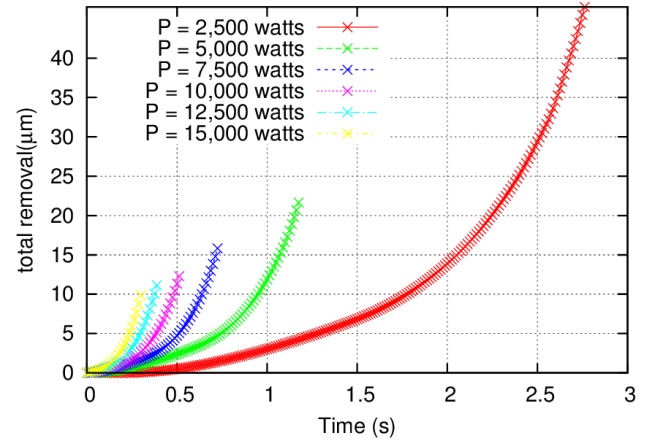

**Fig. 11** Time histories of paint thickness removal for the six laser powers

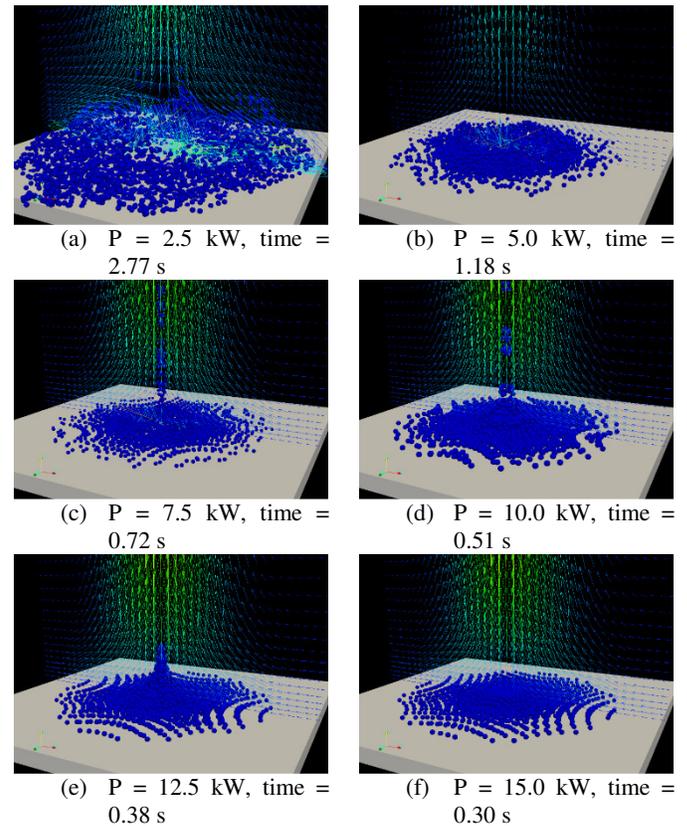

**Fig. 12** Parcel and gaseous flow at the end of the simulation

Figure 11 gives the thickness history of paint removal. It can be seen that the lower the laser intensity is, the more the paint is removed. The removed thicknesses are 40.1, 21.4, 15.9, 12.4, 11.3, and 10.7 μm corresponding to the ascending laser powers of 2.5 kW – 15 kW. Similar to the kinematic energy in the gaseous phase, the trend of thickness reduction here can be explained by the law of energy conservation. In this case, a longer laser heating time can well compensate the energy loss due to the decrement in laser power. As a result, the more energy absorbed, the more paint would be removed by a lower power laser as expected.

Figure 12 shows the behavior of pigments (each dot represents one parcel) after a partial portion of paint is removed by the laser irradiation. It can be seen that all parcels are flowing upward due to laser heating caused natural convection. It can be seen that for the cases with laser power of 7.5 kW, 10.0 kW, 12.5 kW and 15.0 kW, the parcels are at the stage of gathering and moving upward before the simulations are completed. For the cases with power of 2.5 kW and 5.0 kW, it is found that parcels start blowing around by the downward flow due to large number of generated parcels.

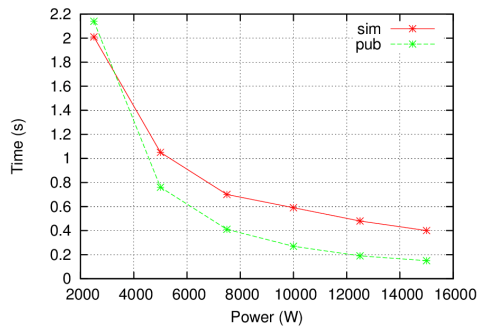

**Fig. 13** Comparison of paint removal between simulation and experiment

This work also attempts to reveal the relationship for the amount of paint removal with laser power and irradiation time. Figure 13 compares the simulation and experiment results of laser power and irradiation time that are needed to remove a portion of 14 µm thick from the paint (for the cases with less than 14 µm paint removal, extrapolation is applied). It should be pointed out that the experiment reported in the reference (Liu and Garmire 1995) might be different from the conditions that are considered in the present study. However, the authors believe that the functional form obtained from that experimental work illustrated a general form of paint-removal rate versus beam intensity. Also due to lack of experimental data, only the result in Reference (Liu and Garmire 1995) is adopted for comparison here. It can be seen from the Fig. 13 where both the trend and the order of magnitude of the simulation results are well in consistence with those experimental data. For the same laser power, the present simulation leads to a longer heating time for the paint removal. The discrepancy between the experimental and simulation results could be caused by a possible reason which is only chemical reaction is taken account for the paint removal in the present simulation model, while, in reality, each of vaporization and chemical reaction could yields a certain amount of material removal. Based on the difference shown in Fig. 13, it may be conjectured that chemical reaction is dominating in the paint removal for lower laser power while vaporization for higher laser power. Further model improvement by including more realistic physical process is suggested.

## 4. CONCLUSIONS

A multi-physics problem that involves compressible gas flow, heat and mass transfer, and chemical reaction is numerically studied for a stainless steel substrate coated with paint which is irradiated by a high energy laser power. Six scenarios with laser powers of 2.5, 5.0, 7.5, 10, 12.5, and 15 kW are considered while the beam diameter is kept at 13.4 mm. A new solver is developed and incorporated into the current OpenFOAM-2.3.0 framework. The numerical simulations are terminated when the maximum temperature of the stainless steel reaches its melting point. The results reveal the effects of laser power on temperature, density, velocity, and species concentration of the gas species around the heated paint. It is found that a higher laser power leads to a shorter simulation time (2.77, 1.18, 0.72, 0.51, 0.38, and 0.30 s), a higher maximum temperature in the paint (1720, 1780, 1820, 1880, 1910, and 1960 K), a lower minimum mass density (0.196, 0.190, 0.185, 0.180, 0.176, and 0.172 kg/m$^3$) and lower velocity (0.895, 1.730, 0.901, 0.866, 0.813, and 0.800 m/s) of gas species, and a smaller amount of paint removal (40.1, 21.4, 15.9, 12.4, 11.3, and 10.7 µm). The variation of species mass concentrations around the heat spot shows how it is affected by the chemical reaction and mass diffusion. It is also found that all the parcels are scattered over the paint surface when the numerical simulations terminate with the current method of calculating initial velocity of parcel. In comparison, the present chemical reaction model predicts the paint removal that is quantitatively consistent with published experimental result. Further model improvement by including more realistic physical process is suggested.

## ACKNOWLEDGEMENTS

Support for this work by the Air Force Research Lab under grant number STTR FA9451-12 is gratefully acknowledged.

## NOMENCLATURE

| | |
|---|---|
| $d$ | Diameter of particles (m) |
| $D_i$ | Mass diffusion coefficient for species i in the mixture (m$^2$/s) |
| $D_{ij}$ | Mass diffusivity coefficient between species $i$ and $j$ |
| $E$ | Activation energy (KJ/mol) |
| **F** | Force (N) |
| **F**$^i_p$ | Force acting on a single particle in a parcel located in the i$^{th}$ cell |
| $g$ | Gravitational acceleration (m/s$^2$) |
| $h$ | Enthalpy(J/kg) |
| **I** | Unit matrix |
| $k_c$ | Chemical reaction rate constant (s$^{-1}$), |
| $k_d$ | Node diffusion coefficient |
| $m$ | Mass of particle |
| N | Number of particles in one parcel |
| N$^i_p$ | Number of particles in a parcel located in the i$^{th}$ cell |
| $p$ | Pressure (Pa) |
| $P$ | Laser power (W) |
| $r_0$ | Radius of laser beam (m) |
| $R$ | Universal gas constant (J/Kg K) |
| $S_h$ | Source term in the energy equation |
| **S**$_U$ | Source term in the momentum equation |
| $t$ | Time (s) |
| $U$ | Velocity of gas (m/s) |
| $V$ | Volume of control volume (m$^3$) |
| $X_i$ | Molar fraction of species i |
| **Greeks** | |
| $\alpha$ | Enthalpy thermal diffusivity (m$^2$/s) |
| $\alpha_{ab}$ | Absorptivity |
| $\Delta Y$ | Moving distance in the y direction (m) |
| $\Delta V$ | Volume of a control volume (m$^3$) |
| $\Delta V_{rem}$ | Volume removed (m$^3$) |
| $\mu$ | Dynamic viscosity(kg/m s) |
| $\rho$ | Mass density (kg/m$^3$) |
| $\psi$ | Compressibility(s$^2$/m$^2$) |
| $\omega_i$ | Mass fraction of species i |
| **Subscript** | |
| $c$ | chemical reaction |
| $bd$ | binder |
| $i$ | index of species |
| $j$ | index of species |
| $p$ | Particle |
| $pg$ | *pigment* |